\documentclass[superscriptaddress,twocolumn,showpacs,aps,pre]{revtex4-1}
\usepackage{amssymb,amsmath}
\usepackage[dvips]{graphicx}

\begin{document}

\title{Strong-randomness infinite-coupling phase in a random quantum spin chain}

\author{Fawaz Hrahsheh}
\affiliation{Department of Physics, Missouri University of Science and Technology, Rolla, MO 65409, USA}

\author{Jos\'e A. Hoyos}
\affiliation{Instituto de F\'{i}sica de S\~ao Carlos, Universidade de S\~ao Paulo,
C.P. 369, S\~ao Carlos, S\~ao Paulo 13560-970, Brazil}

\author{Rajesh Narayanan}
\affiliation{Department of Physics, Indian Institute of Technology Madras, Chennai 600036, India}

\author{Thomas Vojta}
\affiliation{Department of Physics, Missouri University of Science and Technology, Rolla, MO 65409, USA}

\begin{abstract}
We study the ground-state phase diagram of the Ashkin-Teller random quantum spin chain by means of
a generalization of the strong-disorder renormalization group.  In addition to the conventional paramagnetic and
ferromagnetic (Baxter) phases, we find a partially ordered phase characterized by strong randomness
and infinite coupling between the colors. This unusual phase acts, at the same time, as a Griffiths phase for two distinct quantum phase
transitions both of which are of infinite-randomness type.
We also investigate the quantum multi-critical point that separates the two-phase and three-phase regions; and we discuss generalizations
of our results to higher dimensions and other systems.
\end{abstract}

\date{\today}

\maketitle

\section{Introduction}

Random quantum many-particle systems are easiest to understand if both interactions
and disorder are weak. In these cases, the system often behaves analogously to a clean
noninteracting one, with small perturbative corrections.
If, on the other hand, interactions or disorder are
strong, qualitatively new behavior can arise. For instance, repulsive interactions
induce a new phase, the Mott insulator, in systems of lattice bosons or electrons.
Moreover, strong randomness leads to an Anderson insulator in which the quantum wave functions
are localized.

Particularly strong disorder and correlation effects can be expected in the vicinity of
zero-temperature
quantum phase transitions where the fluctuations extend over large length and
time scales. Examples include infinite-randomness criticality \cite{Fisher92,Fisher95},
quantum Griffiths singularities \cite{ThillHuse95,RiegerYoung96} and smeared phase
transitions \cite{Vojta03a} (for recent reviews see, e.g., Refs.\ \cite{Vojta06,Vojta10}).

Disordered quantum spin chains are a paradigmatic class of materials to study these
phenomena, both in theory and in experiment. Theoretically, they have been
attacked by strong-disorder renormalization group (SDRG) methods \cite{MaDasguptaHu79,IgloiMonthus05}
that give asymptotically exact results for a number of one-dimensional systems.
The ground state of the antiferromagnetic spin-$1/2$ random quantum Heisenberg chain
is an exotic random-singlet state controlled by an infinite-randomness renormalization group
fixed point \cite{Fisher94}. Similarly, the ferromagnetic-paramagnetic quantum phase
transition of the random transverse-field Ising chain is of unconventional infinite-randomness
type and accompanied by power-law quantum Griffiths singularities \cite{Fisher95}.
Some of these phenomena have been observed in early experiments on organic crystals
\cite{TheodorouCohen76,TippieClark81} and more recently in MgTiOBO$_3$ \cite{PGFGC07}.

In this paper we investigate
the random quantum Ashkin-Teller model, a prototypical disordered spin chain (or ladder)
that can be understood as two coupled random quantum Ising chains. In addition to quantum
spin systems, versions of the Ashkin-Teller model are used to describe layers
of atoms absorbed on surfaces \cite{Baketal85}, current loops in high-$T_c$ superconductors
\cite{AjiVarma07,*AjiVarma09}, as well as
the elastic response of DNA molecules \cite{ChangWangZheng08}.

\begin{figure}
\includegraphics[width=8.5cm]{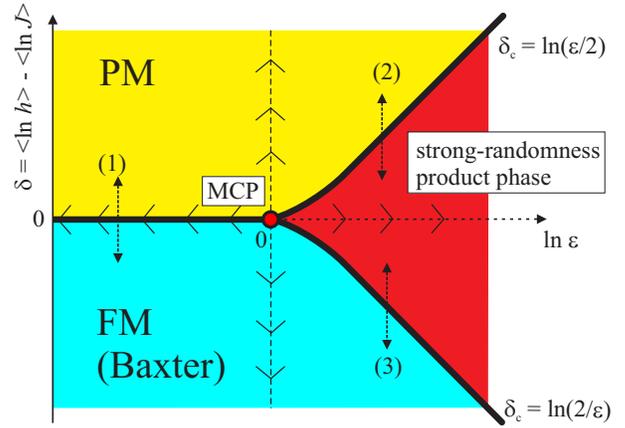}
\caption{(Color online) Schematic ground state phase diagram of the random quantum Ashkin-Teller chain. For $\epsilon<1$, the
paramagnetic and ferromagnetic phases are connected by a direct continuous quantum phase transition. For
$\epsilon>1$, they are separated by a partially ordered ``product'' phase characterized by strong randomness and renormalization
group flow towards infinite coupling. The flow is indicated by arrows on the principal axis, $\delta=0$ and
$\epsilon=1$, of the multicritical point (MCP).}
\label{fig:pd}
\end{figure}

We explore the ground state phase diagram of the random quantum Ashkin-Teller chain by a generalization
of the SDRG technique. In addition to the conventional paramagnetic and
ferromagnetic phases, we identify an unconventional partially ordered phase characterized by finite but strong
randomness and infinite coupling between the two constituent Ising chains (see Fig.\ \ref{fig:pd}).
It plays the role of a ``double-Griffiths'' phase for two separate
quantum phase transitions both of which are of infinite-randomness type. The two-phase region
at weak coupling and the three-phase region at strong coupling are separated by a distinct
infinite-randomness multi-critical point.

The remainder of this paper is organized as follows. We introduce the random quantum Ashkin-Teller model
in Sec.\ \ref{sec:model}. The SDRG method is developed in Sec.\ \ref{sec:SDRG}. Section \ref{sec:PD}
is devoted to the resulting ground state phase diagram and the properties of the quantum phase transitions
between the different phases.   In the concluding section \ref{sec:conclusions},
we discuss generalizations of our results to higher dimensions as well as connections
to other random quantum systems.

\section{Random quantum Ashkin-Teller Model}
\label{sec:model}

The Hamiltonian of the one-dimensional random quantum Ashkin-Teller model is given by \cite{AshkinTeller43,KohmotoNijsKadanoff81,CarlonLajkoIgloi01}
\begin{eqnarray}
 H=&-&\sum_{\alpha=1}^2\sum_{i}{\left ( J_i S_{\alpha,i}^z S_{\alpha,i+1}^z + h_i S_{\alpha,i}^x \right )}\nonumber \\
&-&\sum_{i}{\left (K_i S_{1,i}^z S_{1,i+1}^z S_{2,i}^z S_{2,i+1}^z + g_i S_{1,i}^x S_{2,i}^x\right )}
\label{eq:HAT}
\end{eqnarray}
where  $S^x$ and $S^z$ denote the usual Pauli matrices.
The model can be understood as two identical random transverse-field Ising chains [first line of (\ref{eq:HAT})],
coupled via their energy densities [second line of (\ref{eq:HAT})].
The index $\alpha=1,2$ that distinguishes the two chains is often called the color index.
The strength of the coupling between the colors can be parameterized by the ratios
$\epsilon_{h,i}=g_i/h_i$ and $\epsilon_{J,i}=K_i/J_i$.
Note that the Hamiltonian (\ref{eq:HAT}) is invariant under the duality transformation:
$S_{\alpha,i}^z S_{\alpha,i+1}^z \to \tau_{\alpha,i}^x$,
$S_{\alpha,i}^x \to \tau_{\alpha,i}^z \tau_{\alpha,i+1}^z$,
 $J_i\rightleftarrows h_i$, and $\epsilon_{J,i}\rightleftarrows\epsilon_{h,i}$,
where $\tau^x$ and $\tau^z$ are the dual Pauli operators.

We take the interactions $J_i$ and transverse fields $h_i$ to be
independent random variables. Without loss of generality, we can assume the
$J_i$ and $h_i$ to be positive as possible negative signs can absorbed by local
transformations of the spin variables.
For now, we assume the (bare) coupling strengths to be uniform, $\epsilon_{h,i} = \epsilon_{J,i} =\epsilon_I \ge 0$
\footnote{Even if we assume uniform, nonrandom values of $\epsilon_J$ and $\epsilon_h$,
they will acquire randomness under renormalization.}.
Effects of random $\epsilon$ will be discussed later in Sec.\ \ref{subsec:random_eps}.

The behavior of the random quantum Ashkin-Teller
chain (\ref{eq:HAT}) in the weak-coupling regime, $\epsilon<\epsilon_c=1$,
has been studied in Refs.\
\cite{CarlonLajkoIgloi01,GoswamiSchwabChakravarty08}.
In the following, we therefore focus on the strong-coupling case  $\epsilon \ge \epsilon_c=1$ where these results
do not apply.
For strong coupling, the terms in the second line of (\ref{eq:HAT}) dominate. It is thus
convenient to introduce the product $S_{1,i}^z S_{2,i}^z$ as a new variable.
We define
\begin{eqnarray}
\sigma_i^z &=& S_{1,i}^z S_{2,i}^z~,\label{eq:sigma_z}\\
\eta_i^z &=& S_{1,i}^z~,\\
\sigma_i^z\eta_i^z &=& S_{2,i}^z ~.
\end{eqnarray}
The mapping of the Pauli matrices
$S_{1,i}^x$ and $S_{2,i}^x$ can be easily worked out by exploring their action on a complete set
of basis states in the 4-dimensional single-site Hilbert space. This gives
\begin{eqnarray}
\sigma_i^x &=& S_{2,i}^x~,\\
\eta_i^x &=& S_{1,i}^x S_{2,i}^x~,\\
\sigma_i^x\eta_i^x &=& S_{1,i}^x ~.\label{eq:sigma_x_eta_x}
\end{eqnarray}
Using these transformations, the Hamiltonian (\ref{eq:HAT}) can be rewritten as
\begin{eqnarray}
H = &-& \sum_i (K_i \sigma_i^z\sigma_{i+1}^z + h_i \sigma_i^x) - \sum_i (J_i \eta_i^z\eta_{i+1}^z + g_i \eta_i^x)  \nonumber\\
    &-& \sum_i (J_i\sigma_i^z\sigma_{i+1}^z\eta_i^z\eta_{i+1}^z + h_i \sigma_i^x\eta_i^x)~.
\label{eq:Hprod}
\end{eqnarray}
This form immediately gives an intuitive physical picture of the strong coupling regime $\epsilon \gg 1$ close to
self duality, $h_{\rm typ} \approx J_{\rm typ}$, i.e., close to the horizonal line $\delta=0$ in Fig.\ \ref{fig:pd}.
Here, the typical values of the fields and interactions are defined as $\ln h_{\rm typ} = \langle \ln h \rangle$ and
$\ln J_{\rm typ} = \langle \ln J \rangle$ where $\langle \ldots \rangle$ denotes the disorder average.
The behavior of the product variable $\sigma$ is dominated by the four-spin interactions $K_i$ while the behavior of
the variable $\eta_i$ which traces the original spins is dominated by the two-spin transverse fields $g_i$. Moreover, the coupling
terms between $\sigma$ and $\eta$ are weak. Thus, we expect the system to be in a phase
in which the product variables $\sigma_i$ develop long-range order while the spins remain disordered.

\section{Strong-disorder renormalization group}
\label{sec:SDRG}

To confirm this intuitive picture and to work out the properties of the product phase and its transitions, we now
develop a strong-coupling SDRG. The basic idea of any SDRG consists in identifying the largest local
energy scale and perturbatively integrating out the corresponding high-energy degree of freedom.
As the random quantum Ashkin-Teller model contains four competing local energies $J_i, K_i, h_i, g_i$ rather than
the usual two, we need to generalize the RG scheme by also considering the second-largest
energy in a local cluster.
Details of this calculation are outlined in Appendix \ref{appendix:recursions}.
In the strong-coupling regime, $\epsilon>1$, there are four possible SDRG steps.

\begin{figure}[t,b]
\includegraphics[width=8cm]{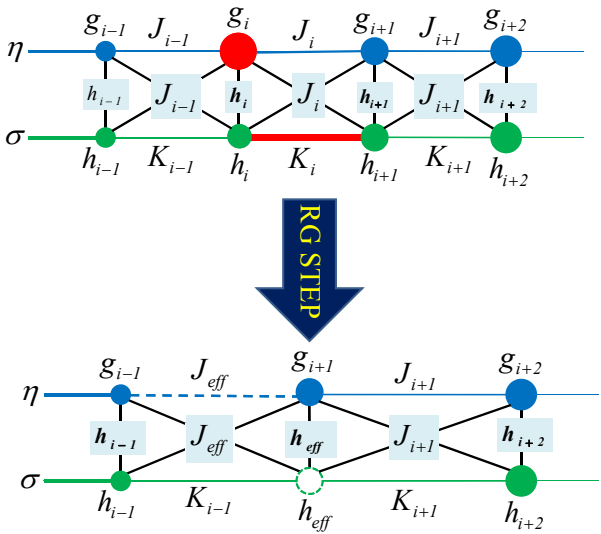}
\caption{(Color online) SDRG steps (a) and (b) decimate a spin variable $\eta_i$ but merge
      two product variables, $\sigma_i$ and $\sigma_{i+1}$, into a new cluster.}
\label{fig:SDRG2}
\end{figure}

(a) If the largest energy in the system is the two-spin field $g_i$, and the second largest energy in the local
cluster of sites $i-1, i$ and $i+1$ is a four-spin interaction, say $K_i$, the SDRG step decimates the variable
$\eta_i$ but merges $\sigma_i$ and $\sigma_{i+1}$ to a new cluster $\tilde \sigma$.
The unperturbed Hamiltonian for this SDRG step reads $H_0 = -K_i \sigma_i^z \sigma_{i+1}^z -g_i \eta_i^x$.
We now keep only the ground state of $H_0$ and treat all other terms that contain $\sigma_i, \sigma_{i+1}$
or $\eta_i$ in second order perturbation theory.
 The resulting Hamiltonian has the same form as (\ref{eq:Hprod}) with one fewer site and renormalized
energies arranged as shown in Fig.\ \ref{fig:SDRG2}.
\begin{eqnarray}
\tilde J_{\rm eff} = \frac {2J_{i-1}J_i}{g_i}, \qquad \tilde h_{\rm eff} = \frac {2h_i h_{i+1}}{K_i}
\label{eq:SDRG_AT}
\end{eqnarray}
As $\tilde J_{\rm eff}$ and $\tilde h_{\rm eff}$ are renormalized downward while all remaining
$K_i$ and $g_i$ are unchanged, the coupling strengths  $\epsilon_{h,i}$ and  $\epsilon_{J,i}$
increase under renormalization.

(b) The same SDRG step is carried out if the largest energy is the four-spin interaction $K_i$, and the
second-largest energy in the cluster of sites $i$ and $i+1$ is a two-spin field, say $g_i$.

(c) If the largest energy in the system is the two-spin field $g_i$, and the second largest energy in the local
cluster of sites $i-1, i$ and $i+1$ is the field $h_i$, both $\sigma_i$ and $\eta_i$ are decimated. This is equivalent
to decimating both original spins $S_{1,i}$ and $S_{2,i}$ and leads to the recursion relations
\begin{eqnarray}
\tilde K_{\rm eff} = \frac {K_{i-1} K_i}{2 h_i}, \qquad \tilde J_{\rm eff} = \frac {J_{i-1} J_{i}}{g_i+h_i}
\label{eq:SDRG_site}
\end{eqnarray}
for the interaction energies that emerge between sites $i-1$ and $i+1$ in the renormalized chain.
This implies that the renormalized coupling strength $\tilde \epsilon_{J} = \epsilon_{J,i-1}\epsilon_{J,i}(1+\epsilon_{h,i})/2$
increases under renormalization (as we are interested in the strong-coupling regime $\epsilon > \epsilon_c=1$).

(d) Finally, if the largest energy is the four-spin interaction $K_i$, and the second-largest energy associated with
the sites $i$ and $i+1$ is the interaction $J_i$, clusters are formed from $\sigma_i$ and $\sigma_{i+1}$
as well as $\eta_i$ and $\eta_{i+1}$. This is equivalent to forming clusters of both original spin variables,
$S_{1,i}$ and $S_{1,i+1}$ as well as $S_{2,i}$ and $S_{2,i+1}$. The resulting recursions for the transverse
field $\tilde h_{\rm eff}$ and two-spin field $\tilde g_{\rm eff}$ acting on these clusters read
\begin{eqnarray}
\tilde g_{\rm eff} = \frac {g_{i} g_{i+1}}{2 J_i}, \qquad \tilde h_{\rm eff} = \frac {h_{i} h_{i+1}}{K_i+J_i}~.
\label{eq:SDRG_bond}
\end{eqnarray}
The renormalized coupling strength $\tilde \epsilon_{h} = \epsilon_{h,i}\epsilon_{h,i+1}(1+\epsilon_{J,i})/2$
increases under renormalization.

The SDRG steps (a) to (d) are now iterated. As a result, the maximum local energy $\Omega$ in the system gradually
decreases from its initial (bare) value $\Omega_I$.

\section{Phase diagram and phase transitions}
\label{sec:PD}

\subsection{Double Griffiths phase}

Based on the SDRG recursions (\ref{eq:SDRG_AT}) to (\ref{eq:SDRG_bond}), the phase diagram of the random quantum
Ashkin-Teller model shown in Fig.\ \ref{fig:pd} is easily understood. Let us start by recalling that in the weak-coupling
regime, $\epsilon<1$, the local coupling strengths $\epsilon_{h,i}$ and $\epsilon_{J,i}$ decrease without limit
under renormalization \cite{CarlonLajkoIgloi01,GoswamiSchwabChakravarty08}. This
implies that the two Ising chains that make up the Ashkin-Teller model decouple in the low-energy limit.
Our system thus behaves analogously to the random  transverse-field
Ising chain \cite{Fisher92,Fisher95}: A paramagnetic phase at large transverse fields $h_i$ and a
ferromagnetic phase at large interactions $J_i$ are directly connected  by an infinite-randomness critical
point at $\delta=\ln h_{\rm typ} -\ln J_{\rm typ}=0$  (transition 1 in Fig.\ \ref{fig:pd})
\footnote{The position of this phase boundary is fixed by the self-duality of the Hamiltonian.}.

To understand the strong-coupling regime $\epsilon>1$, let us first focus on the
self-duality line $\delta=\ln h_{\rm typ} -\ln J_{\rm typ}=0$. If the bare $\epsilon_I$ is just slightly
above 1, most of the recursions will initially be site and bond decimations [types (c) and (d)].
In these steps, the local coupling strengths $\epsilon_{h,i}$ and $\epsilon_{J,i}$ rapidly increase.
When they become larger than the widths of the $J$ and $h$ distributions, the character of the SDRG changes.
Now, most steps are ``mixed steps'' of types (a) and (b). As a result, the product variable $\sigma$
forms larger and larger clusters while the spin variable $\eta$ is decimated.

The system
is thus in a ``double Griffiths phase:'' The $\sigma$-part of the Hamiltonian behaves analogously to an ordered
Griffiths phase while the $\eta$-part behaves as in a disordered Griffiths phase.
This double Griffiths phase has a nonzero product order parameter or polarization $M_p = \sum_i  \sigma_i^z $ while the spin variable
$\eta_i^z$ (and thus $S_{1,i}^z$ and $S_{2,i}^z$) remains disordered, $M=\sum_i  \eta_i =0$.
Note that this behavior is valid not just on the self-duality line, $\delta=0$, but also in its vicinity
because the RG flow of each of the variables $\sigma$ and $\eta$ is dominated by a single term in the
Hamiltonian and does not rely on the balance between interactions and transverse fields. Thus, we have indeed
discovered a bulk phase rather than a special line in the phase diagram. Moreover, as  the
$\epsilon_{h,i}$ and $\epsilon_{J,i}$ flow to infinity, the analysis is asymptotically exact.

To find the extensions of the partially ordered double Griffiths phase we need to locate its transitions to the conventional
paramagnetic and ferromagnetic phases. Looking at the first sum in the Hamiltonian (\ref{eq:Hprod}),
it is clear that the long-range order of the product variable will be destroyed if we raise $\delta=\ln h_{\rm typ} -\ln J_{\rm typ}$ until
the transverse fields $h_i$ compete with the four-spin interactions $K_i$. This leads to a competition between
the SDRG steps (a) and (c). From comparing the $h$ recursion in (\ref{eq:SDRG_AT}) with the $K$ recursion in (\ref{eq:SDRG_site}),
we conclude that the phase boundary between the double Griffiths phase and the paramagnetic phase
(transition 2 in Fig.\ \ref{fig:pd}) is located at $K_{\rm typ} = 2 h_{\rm typ}$ or equivalently $\delta_c=\ln(\epsilon/2)$ in the limit
of large $\epsilon$. Moreover, the transition is governed by an infinite-randomness fixed point in the random transverse-field Ising universality
class. The phase boundary to the ferromagnetic phase (transition 3 in Fig.\ \ref{fig:pd}) can be found analogously.
For large $\epsilon$, it is located at $2 J_{\rm typ} = g_{\rm typ}$ or equivalently $\delta_c = \ln(2/\epsilon)$ in agreement with the
self-duality of the Hamiltonian.

\begin{figure}[t,b]
\includegraphics[width=7.5cm]{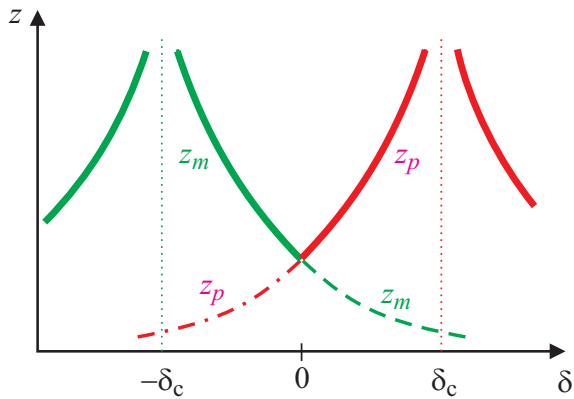}
\caption{(Color online) Schematic of the Griffiths dynamical exponent $z$ as a function of
 $\delta$ for fixed $\epsilon>1$. Rare large magnetization clusters lead to Griffiths singularities associated with the exponent $z_m$
 (green dashed line) while
 the singularities stemming from rare polarization clusters are associated with $z_p$ (red dash-dotted line).
 Thermal quantities are dominated by the larger of the two exponents $z=\max(z_m,z_p)$  which features
 non-monotonous behavior (thick solid line).}
 \label{fig:z}
\end{figure}

The thermodynamics of the double Griffiths phase is highly unusual. It can be found in the usual way, i.e., by including
conjugate fields into the SDRG.
Each of the two order parameters, the magnetization
$M=\sum_i \eta_i$ and the polarization $M_p = \sum_i \sigma_i^z$, displays power-law quantum Griffiths singularities
controlled by different Griffiths dynamical exponents $z_m$ and $z_p$, respectively, that vary non-universally with $\epsilon$ and $\delta$.
The exponent $z_m$ diverges at the transition to the ferromagnetic phase while $z_p$ diverges at the transition to the
paramagnetic phase. Duality imposes the relation $z_p(\epsilon,\delta)=z_m(\epsilon,-\delta)$. Thermal quantities such as
the entropy and the specific heat pick up contributions from both order parameters. Their Griffiths dynamical exponent
$z = \max(z_m,z_p)$ thus displays an interesting non-monotonous dependence on $\delta$, as sketched in Fig.\
\ref{fig:z}.

\subsection{Multicritical point}
\label{sec:MCP}

Finally, we consider the infinite-randomness multicritical point (MCP) located at $\epsilon=1, \delta=0$.
It has two independent unstable directions, the lines $\delta=0$ and $\epsilon=1$.
On the line $\epsilon=1$ that separates the weak-coupling and strong-coupling regimes,
all SDRG steps are site and bond decimations (types (c) or (d)). The recursions
(\ref{eq:SDRG_site}) and (\ref{eq:SDRG_bond}) reduce to the well-known transverse-field
Ising forms $\tilde J_{\rm eff} = {J_{i-1} J_{i}}/{(2h_i)}$
and $h_{\rm eff} = {h_{i} h_{i+1}}/({2 J_i)}$ \footnote{The extra factors of
2 in the denominator are irrelevant in the low-energy limit.}. The SDRG flow of the
$J$ and $h$ distributions on the line $\epsilon=1$ is thus identical to the
corresponding flow of the random-transverse-field Ising chain.
We emphasize, however, that although the SDRG flow of the $J$ and $h$  distributions
at $\epsilon=1$ is identical to the weak-coupling regime $\epsilon<1$, the fixed-point Hamiltonian differs
because the two Ising chains  that make up the Ashkin-Teller model do not decouple.

The flow along the line $\epsilon=1$ can be
characterized by the following critical singularities: correlation length $\xi \sim |\delta|^{-\nu}$,
magnetization $M\sim |\delta|^\beta$,  and correlation time $\ln \xi_t\sim \xi^\psi$ with exponents
\begin{equation}
\nu=2, \quad \beta=2-(1+\sqrt{5})/2=0.382,\quad \psi=1/2
\label{eq:MCP_delta}
\end{equation}
In contrast, the SDRG flow on the self-duality line $\delta=0$ for $\epsilon>1$ close to the MCP is determined
by the evolution of $\epsilon$ under repeated site and bond decimations (steps (c) and (d)).
It can be worked out (see Appendix \ref{appendix:MCP}) by including $\ln(\epsilon)$ as an auxiliary variable
in the SDRG flow of the $J$ and $h$  distributions. We find different critical singularities
$\xi \sim (\epsilon-1)^{-\nu_\epsilon}$ and $M_p \sim (\epsilon-1)^{\beta_\epsilon}$ with
exponents
\begin{equation}
\nu_\epsilon=\frac{8}{1+\sqrt{7}}= 2.194~, \quad \beta_\epsilon=\frac{6-2\sqrt{5}}{1+\sqrt{7}}=0.419~.
\label{eq:MCP_eps}
\end{equation}
The tunneling exponent $\psi$ remains 1/2. Combining these results to write a scaling
form of the polarization gives
\begin{equation}
M_p(\delta,\epsilon-1) = b^{-\beta/\nu} M_p(\delta b^{1/\nu}, (\epsilon-1) b^{1/\nu_\epsilon})
\label{eq:Mp_scaling}
\end{equation}
where $b$ is an arbitrary scale factor. The phase transition between the partially ordered and paramagnetic phases
corresponds to a singularity of $M_p$ for $\delta>0$ and $\epsilon>1$. Using (\ref{eq:Mp_scaling}), we find
that the phase boundary behaves as
\begin{equation}
\delta_c \sim (\epsilon-1)^{\nu_\epsilon/\nu} = (\epsilon-1)^{4/(1+\sqrt{7})} =(\epsilon-1)^{1.097}
\label{eq:phaseboundary}
\end{equation}
sufficiently close to the multicritical point. The phase boundary to the ferromagnetic
phase can be found analogously.

\subsection{Random coupling strength $\epsilon$}
\label{subsec:random_eps}

So far, we have considered systems in which the (bare) coupling strengths are uniform
$\epsilon_{J,i}=\epsilon_{h,i}=\epsilon_I$. In the present section, we discuss what changes
for random coupling strengths.

If \emph{all} $\epsilon_{J,i}$ and $\epsilon_{h,i}$ are below the multicritical value of 1,
the renormalized values $\tilde\epsilon$ are also smaller than 1 and decrease under
renormalization. Thus, the two Ising chains that make up the Ashkin-Teller model decouple
in the low-energy limit, just as in the case of uniform bare $\epsilon$.
Conversely, if
\emph{all} $\epsilon_{J,i}$ and $\epsilon_{h,i}$ are above the multicritical value of 1,
the renormalized values $\tilde\epsilon$ are also larger than 1 and increase under
renormalization. The system thus flows to the strong coupling region, also
just as in the case of uniform bare $\epsilon$.
Consequently, none of our results change in these two cases, except for
unimportant modifications of nonuniversal quantities.  This also implies
that the three bulk phases shown in Fig.\ \ref{fig:pd} are stable against weak
randomness in $\epsilon$. The same holds for the phase transitions (1), (2) and
(3) sufficiently far away from the multicritical point discussed in Sec.\
\ref{sec:MCP}.

In contrast, the multicritical point at $\delta=0,\epsilon=1$ itself is unstable
against weak disorder in the $\epsilon_{J,i}$ and $\epsilon_{h,i}$. To show this, we
analyze how the width of a narrow $\epsilon$-distribution around $\epsilon=1$ flows under repeated SDRG
site and bond decimations. By including $\ln \epsilon$ as an auxiliary variable in the SDRG and
using the methods of Ref.\ \cite{Fisher94}, we find
\begin{equation}
\sigma_{\ln\epsilon} \sim \Gamma^{\phi^{(\rm sym)}_{1/2}}~, \qquad \phi^{(\rm sym)}_{1/2} = \frac {1+\sqrt{6}} 4~.
\label{eq:width1}
\end{equation}
(Note that we need to consider the flow of a \emph{symmetrically} distributed auxiliary variable;
the exponent is therefore denoted as $\phi^{(\rm sym)}$.)
This means that a narrow bare distribution broadens under the SDRG, destabilizing the
uniform-$\epsilon$ multicritical point of Sec.\ \ref{sec:MCP}.

\begin{figure}
\includegraphics[width=8.5cm,clip]{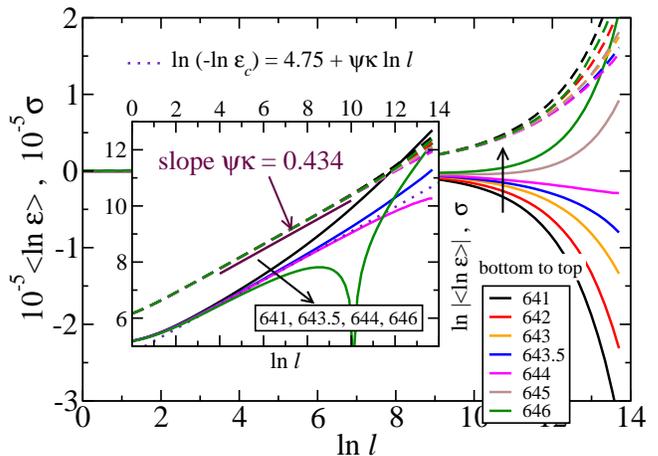}
\caption{(Color online) Average value $\langle \ln\epsilon\rangle$ (solid lines) and standard deviation $\sigma_{\ln\epsilon}$
(dashed lines) of the coupling strength $\ln\epsilon$ versus the SDRG length scale $\ln \ell$ for different values of the tuning parameter
$\ln\epsilon_{\textrm{max}}=641 \ldots 646$. The inset shows  $\ln|\langle \ln\epsilon\rangle|$ and $\ln \sigma_{\ln\epsilon}$ for
selected curves, giving a multicritical value of $\ln \epsilon_c=643.75$ }
\label{fig:MCP_kappa}
\end{figure}

\begin{figure}
\includegraphics[width=8.5cm,clip]{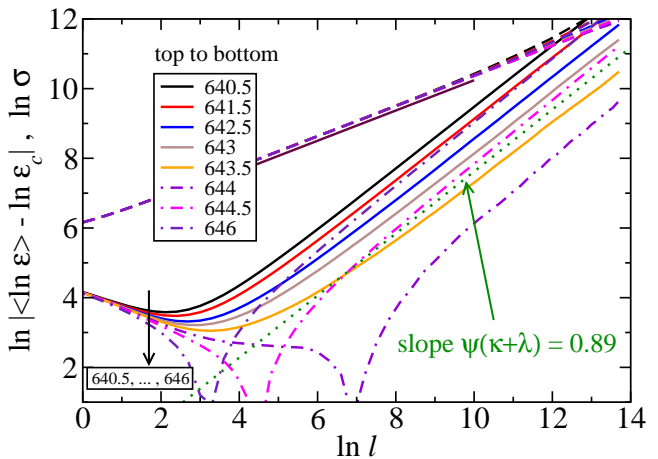}
\caption{(Color online) Distance $\ln|\langle \ln \epsilon \rangle -\ln \epsilon_c|$ from the multicritical point
versus the SDRG length scale $\ln \ell$ for different values of the tuning parameter
$\ln\epsilon_{\textrm{max}}=640.5 \ldots 646$.}
\label{fig:MCP_lambda}
\end{figure}

We have not found an analytic solution of the multicritical behavior in
the case of random $\epsilon_{J,i}$ and $\epsilon_{h,i}$. Instead, we implement
the SDRG numerically. We study systems with up to $5\times 10^8$ sites. To place
the system on the self-duality line $\delta=\ln h_{\rm typ} -\ln J_{\rm typ}=0$, we employ identical
power-law distributions $P_I(J)=J^{-1+1/w}/w$ and $R_I(h)=h^{-1+1/w}/w$ for the interactions
and transverse fields, with $w$ being a measure of the disorder. The coupling strengths
$\ln\epsilon$ are drawn from a box distribution between $\ln\epsilon_{\textrm{min}}$ and
$\ln\epsilon_{\textrm{max}}$.
The results of a strongly disordered ($w=2000$) example system are summarized in
Figs.\ \ref{fig:MCP_kappa} and \ref{fig:MCP_lambda}. We fix $\ln\epsilon_{\textrm{min}}=-1000$
and tune the multicritical point by varying $\ln\epsilon_{\textrm{max}}$.
The data are averages over 50 different chains of $5\times10^7$ sites each.
Fig.\ \ref{fig:MCP_kappa} shows how the average $\langle \ln \epsilon \rangle$
and standard deviation $\sigma_{\ln \epsilon}$ of the coupling strength
evolve under the SDRG.
From the inset, we determine the multicritical point to be located between $\ln\epsilon_{\textrm{max}}=643.5$
and 644. Moreover, $\sigma_{\ln\epsilon}$ increases as $\ell^{\psi\kappa}$ with $\psi\kappa=0.434(3)$ with the
SDRG length scale. Here, the number in brackets gives the error of the last digit. This error is mostly due
to the uncertainty in precisely locating the multicritical point. The statistical error is much smaller.
As the tunneling exponent remains at $\psi=1/2$, this implies
\begin{equation}
\sigma_{\ln\epsilon} \sim \Gamma^{\kappa}~, \qquad \kappa=0.868(6)
\label{eq:width2}
\end{equation}
The value of the exponent $\kappa$
\footnote{We call this exponent $\kappa$ rather than $\psi$ (as was done in Ref.\ \cite{Fisher94})
to avoid confusion with the tunneling exponent.}
fulfills the constraint $\kappa<1$ derived by Fisher \cite{Fisher94}.
Interestingly, it is not very different from the value $\phi^{(\rm sym)}_{1/2}\approx 0.8624$ that describes
the initial increase of $\sigma_{\ln\epsilon}$ near the uniform-$\epsilon$ multicritical point.

In Fig.\ \ref{fig:MCP_lambda}, we study how the distance $|\langle \ln \epsilon \rangle -\ln \epsilon_c|$
from the multicritical point increases with SDRG length scale $\ell$ in the regime
$|\langle \ln \epsilon \rangle -\ln \epsilon_c| < \sigma_{\ln \epsilon}$.
We find $|\langle \ln \epsilon \rangle -\ln \epsilon_c| \sim \ell^{\psi(\kappa+\lambda)}$
with $\psi(\kappa+\lambda)=0.89(2)$
\footnote{Our exponent $\lambda$ is equivalent to the exponent $\lambda$ used in Ref.\ \cite{Fisher94}
to describe the scaling of the average anisotropy in an XXZ spin chain.}.
Expressed in terms of $\Gamma$, this means
\begin{equation}
|\langle \ln \epsilon \rangle -\ln \epsilon_c| \sim \Gamma^{\kappa+\lambda}~, \qquad \kappa+\lambda=1.78(4)~.
\label{eq:mean}
\end{equation}
Again, the error is mostly due to uncertainties in the location of the multicritical point as well
as the fit range.

We have performed analogous calculations for a number of different parameter sets. For the weaker disorder case
of $w=3$ and $\ln\epsilon_{\textrm{min}}=-3$, the multicritical point is located at $\ln\epsilon_{\textrm{max}}
\approx 2.424$. In this case, our analysis of 180 chains of $5\times 10^8$ sites
gives the same value as above, $\kappa\psi=0.434(3)$.
The exponent $\psi(\kappa+\lambda)$ is somewhat harder to determine in the weak-disorder case because the available fit range
becomes very narrow. We find $\psi(\kappa+\lambda)=0.88(4)$ in agreement with the strong-disorder value.
Further calculations for even weaker disorder and shorter chains (between $10^6$ and $5\times 10^7$ sites) are less precise but compatible with the values
given above.

Once $|\langle \ln \epsilon \rangle -\ln \epsilon_c| > \sigma_{\ln \epsilon}$,
almost all $\epsilon$ are on the same side of the multicritical point. The further
analysis therefore follows the steps outlined in Appendix \ref{appendix:MCP}.
The resulting multicritical behavior along the self-duality line on the strong-coupling side of the
MCP is characterized by the power laws
$\xi \sim (\langle \ln \epsilon \rangle -\ln \epsilon_c)^{-\nu_\epsilon}$ and $M_p \sim (\langle \ln \epsilon \rangle -\ln \epsilon_c)^{\beta_\epsilon}$ with
exponents
\begin{equation}
\nu_\epsilon=\frac{4-2\kappa}{\lambda} = 2.48(15) , ~~ \beta_\epsilon=\frac{2-\kappa}{\lambda}(2-\phi_0)=0.474(20).
\label{eq:MCP_eps_random}
\end{equation}
The shape of the phase boundary close to the multicritical point can be found
as in Sec.\ \ref{sec:MCP} yielding $\delta_c \sim (\langle \ln\epsilon \rangle -\ln\epsilon_c)^{1.24}$.

\section{Discussion and Conclusions}
\label{sec:conclusions}

In summary, we have investigated the ground state phase diagram of the random quantum Ashkin-Teller
spin chain. The topology of the phase diagram, shown in Fig.\ \ref{fig:pd}, is analogous to that of the clean quantum Ashkin-Teller model
(see, e.g., Ref.\ \cite{IgloiSolyom84}). However, the properties of the phases and phase transitions are
different. In addition to the usual paramagnetic and ferromagnetic phases, we have identified a partially ordered phase
characterized by strong randomness and infinite coupling between the colors. This phase acts as a Griffiths phase for two
distinct quantum phase transitions leading to an unconventional non-monotonic variation of the Griffiths dynamical exponent
throughout the phase.

We now turn our attention to the phases boundaries between the three phases.
The direct transition at weak intercolor coupling between the paramagnetic and ferromagnetic (Baxter) phases
(transition (i) in Fig.\ \ref{fig:pd}) is in the infinite-randomness
universality class of the random transverse-field Ising chain, as was already found in Refs.\
\cite{CarlonLajkoIgloi01,GoswamiSchwabChakravarty08}. In contrast, the corresponding phase boundary in the
clean quantum Ashkin-Teller chain shows an unusual line of fixed points with continuously varying exponents
\cite{KohmotoNijsKadanoff81,Baxter_book82}.
The quantum phase transitions separating the partially ordered phase from the paramagnetic and
ferromagnetic phases (transitions (ii) and (iii)) are also of infinite-randomness type and in the
universality class of the random transverse-field Ising chain, while they are in the (1+1)-dimensional
Ising universality class in the clean model.

We have also studied the quantum
multicritical point separating the two-phase and three-phase regions.
It is in one of two different universality classes (both of infinite-randomness type), depending on whether the
intercolor coupling strengths $\epsilon$ are uniform or random. This differs from the infinite-order
multicritical behavior seen in the clean case \cite{KohmotoNijsKadanoff81,Baxter_book82}.

Generalizations of the Ashkin-Teller Hamiltonian (\ref{eq:HAT}) to $n>2$ colors have recently reattracted
considerable attention because they have been used to analyze the fate of first-order
quantum phase transitions under the influence of disorder
\cite{GoswamiSchwabChakravarty08,GreenblattAizenmanLebowitz09,HrahshehHoyosVojta12}.
Interestingly, for $n>4$ colors, the paramagnetic and ferromagnetic phases meet directly at the self-dual line
$\langle \ln h \rangle = \langle \ln J \rangle$ for all coupling strength $\epsilon\ge 0$. Thus an analog to the
partially ordered strong-coupling phase does not exist. For three and four colors, this question is not yet solved
to the best of our knowledge.

The random quantum Ashkin-Teller chain (\ref{eq:HAT}) with $N$ sites can be mapped onto a random XXZ quantum spin chain
with $2N$ sites \cite{AlcarazBarberBatchelor88}. Under this mapping,
the transverse fields $h_i$ in the Ashkin-Teller model map onto the even bonds of the XXZ chain
while the interactions $J_i$ map onto the odd bonds. The coupling strengths $\epsilon_{h,i}$  and $\epsilon_{J,i}$
map onto the local anisotropies of the XXZ chain. Importantly, the mapping is nonlocal as it involves (semi-infinite)
chains of operators. Thus, although the energy spectra of the Ashkin-Teller model and the XXZ chain are analogous,
their order parameters are not directly related. This explains, for example, why the correlation length exponent $\nu_\epsilon$ given in (\ref{eq:MCP_eps})
takes the same value as the exponent that describes the effects of weak anisotropy about the Heisenberg fixed point of the XXZ chain \cite{Fisher94}.
In contrast, our order parameter exponent $\beta_\epsilon$ does not have a direct counterpart in the XXZ chain.

Our study has focused on one space dimension. Let us briefly comment on  random quantum
Ashkin-Teller models in higher dimensions. The crucial step in our  understanding of the strong-coupling regime
was the transformation defined in eqs.\ (\ref{eq:sigma_z}--\ref{eq:sigma_x_eta_x}) from the original spins to
the product variable. This transformation is purely local and can be performed in the same way in any space dimension.
We therefore believe that the basic features of the phase diagram in higher dimensions will be similar to the
one dimensional case. In particular, for small $\epsilon$, we expect a direct transition between the ferromagnetic and
paramagnetic phases while a partially ordered product phase is expected to intervene between them for large $\epsilon$.
Obtaining quantitative results in higher dimensions will be significantly more complicated than in one dimension.
First, the Hamiltonian is not self-dual in $d>1$, thus the phase diagram is not symmetric under the exchange of transverse
fields and interactions. Second, the SDRG can only be implemented numerically in $d>1$ because the decimation steps
change the topology of the lattice. This work remains as a task for the future.

\section*{Acknowledgements}

This work was supported by the NSF under Grant Nos.\ DMR-1205803
and PHYS-1066293, by Simons Foundation, by FAPESP under Grant No.\ 2013/09850-7, and by CNPq under Grant
Nos.\ 590093/2011-8 and 305261/2012-6. R.N. acknowledges the hospitality of the Physics Department of
Missouri S\&T where this works was initiated. J.H. and T.V. acknowledge the hospitality of the
Aspen Center for Physics.

\appendix
\section{SDRG recursion relations}
\label{appendix:recursions}

A single step of the SDRG consists in identifying the largest local energy scale in the
Hamiltonian and perturbatively integrating out the corresponding high-energy excitations.
This is done using the projection technique described, e.g., in Ref.\ \cite{Auerbach98}.
The Hilbert space is divided into a low-energy subspace and a high-energy subspace. Any wave
function $\psi$ can be decomposed as $\psi = \psi_1 +\psi_2$ with $\psi_1$ in the low-energy
subspace and $\psi_2$ in the high-energy subspace. This allows us to write the Schroedinger
equation in matrix form
\begin{equation}
\left( \begin{matrix} H_{11} & H_{12}\\H_{21} & H_{22} \end{matrix}\right) \left ( \begin{matrix} \psi_1 \\ \psi_2 \end{matrix} \right)
 = E \left ( \begin{matrix} \psi_1 \\ \psi_2 \end{matrix} \right)
\label{eq:matrix}
\end{equation}
with $H_{ij} = P_i H P_j$. Here, $P_1$ and $P_2$ project on the low-energy and high-energy subspaces,
respectively. Eliminating $\psi_2$ from these two coupled equations gives
$H_{11} \psi_1 + H_{12} (E-H_{22})^{-1} H_{21} \psi_1 = E \psi_1$.
Thus, the effective Hamiltonian in the low-energy Hilbert space is
\begin{equation}
H_{\rm eff}= H_{11} + H_{12} (E-H_{22})^{-1} H_{21}~.
\label{eq:projected}
\end{equation}
The second term can now be expanded  in inverse powers of the large local energy scale.

The quantum Ashkin-Teller Hamiltonian has four competing local energy scales, viz.,
$J_i, K_i, h_i$, and $g_i$ rather than two. We therefore generalize the usual SDRG scheme by
considering the largest and second-largest energies in a local cluster to define the
SDRG step. In the strong-coupling regime, $\epsilon>1$, the largest local energy is always
either a four-spin interaction or a two-spin field. In total, there are four possible steps.

(a) The largest local energy is a two-spin field $g_i$. The second largest energy in the
three-site cluster of sites $i-1$, $i$, $i+1$ is a four-spin interaction, either $K_{i-1}$ or $K_i$.
Let us assume that it is $K_i$ for definiteness. In this case, the low-energy Hilbert space is spanned by states for which
$(\eta_i, \sigma_i, \sigma_{i+1})=(\to,\uparrow,\uparrow)$ or $(\to,\downarrow,\downarrow)$.
$H_{11}$ and $H_{22}$ contain all terms in the Hamiltonian that do not flip the spins $\eta_i, \sigma_i, \sigma_{i+1}$;
their leading terms are $-K_i \sigma_i^z \sigma_{i+1}^z - g_i \eta_i^z$. All terms that flip at least one of the variables
$\eta_i, \sigma_i, \sigma_{i+1}$ are contained in $H_{12}$ and $H_{21}$. Specifically,
\begin{eqnarray}
H_{12}= P_1 [ &-&J_{i-1}\eta_{i-1}^z\eta_i^z   -J_{i}\eta_{i}^z\eta_{i+1}^z -h_i\sigma_i^x -h_{i+1}\sigma_{i+1}^x  \nonumber\\
&-&h_i\sigma_i^x\eta_i^x -h_{i+1}\sigma_{i+1}^x\eta_{i+1}^x \nonumber\\
&-&J_{i-1}\sigma_{i-1}^z\sigma_i^z\eta_{i-1}^z\eta_i^z -J_{i}\sigma_{i}^z\sigma_{i+1}^z\eta_{i}^z\eta_{i+1}^z]P_2 ~.~~~
\label{eq:H12a}
\end{eqnarray}
$H_{21}$ takes the same form but with $P_1$ and $P_2$ exchanged.
We now insert $H_{12}$ and $H_{21}$ into (\ref{eq:projected}) and approximate the denominator $E-H_{22}$ by
$-2g_i$ or $-2K_i$ depending on which of the spins $(\eta_i, \sigma_i, \sigma_{i+1})$ is flipped. The resulting
effective Hamiltonian has the same form (\ref{eq:Hprod}) as the initial one, but with one fewer site. The arrangement
of the renormalized energies  $\tilde J_{\rm eff}$ and $\tilde h_{\rm eff}$ between the remaining sites
is shown in Fig.\ \ref{fig:SDRG2} and their values are given in (\ref{eq:SDRG_AT}).

(b) Exactly the same SDRG step is carried out if the largest local energy is the four-spin interaction $K_i$, and the
second-largest energy in the two-site cluster of sites $i$ and $i+1$ is a two-spin field, either $g_i$ or $g_{i+1}$.

Steps (a) and (b) are the dominant SDRG steps for $\epsilon \gg 1$. More precisely, most steps are of types (a) or (b)
if $\epsilon$ is larger than the width of the $h$ and $J$ distributions (on a logarithmic scale). In the opposite case,
strong disorder and not too large $\epsilon$, most SDRG steps are site and bond decimations of types (c) and (d).

(c) The largest energy in the system is the two-spin field $g_i$, and the second largest energy in the local
cluster of sites $i-1, i$ and $i+1$ is the field $h_i$. In this case, the low-energy Hilbert space is spanned by all states
having $(\eta_i,\sigma_i)=(\to \to)$. $H_{11}$ and $H_{22}$ contain all terms in the Hamiltonian that do not flip
$\eta_i$ and $\sigma_i$, with the leading terms being $-g_i\eta_i^x-h_i\sigma_i^x-h_i\eta_i^x\sigma_i^x$. All terms that flip
$\eta_i$ and/or $\sigma_i$ are part of $H_{12}$ and $H_{21}$. Specifically,
\begin{eqnarray}
H_{12}= P_1 [&-&K_{i-1} \sigma_{i-1}^z\sigma_i^z -K_i\sigma_i^z\sigma_{i+1}^z \nonumber\\
             &-&J_{i-1} \eta_{i-1}^z\eta_i^z - J_i\eta_i^z\eta_{i+1}^z \nonumber \\
             &-&J_{i-1} \eta_{i-1}^z\eta_i^z \sigma_{i-1}^z\sigma_i^z - J_i\eta_i^z\eta_{i+1}^z \sigma_i^z\sigma_{i+1}^z ]P_2~.~~~
\label{eq:H12c}
\end{eqnarray}
and $H_{21}$ takes the same form but with $P_1$ and $P_2$ exchanged. After inserting this into (\ref{eq:projected}) and
approximating the denominator $E-H_{22}$ by $-2g_i-2h_i$ or $-4h_i$ depending on which spins are flipped, site $i$ is
eliminated (i.e., both $\sigma_i$ and $\eta_i$ are decimated). The effective interaction energies between the neighboring sites
$i-1$ and $i+1$ are given in (\ref{eq:SDRG_site}).

(d) The largest local energy is a four-spin interaction $K_i$, and the second largest energy in the cluster consisting of
sites $i$ and $i+1$ is the interaction $J_i$. The low-energy Hilbert space is spanned by states having $(\eta_i,\eta_{i+1},\sigma_i,\sigma_{i+1})
= (\uparrow\uparrow\uparrow\uparrow)$ or $(\uparrow\uparrow\downarrow\downarrow)$ or $(\downarrow\downarrow\uparrow\uparrow)$  or
$(\downarrow\downarrow\downarrow\downarrow)$. After projection into the low-energy Hilbert space, the two sites
  $i$ and $i+1$ can thus be represented by a single site with variables $\sigma_{\rm eff}$ and $\eta_{\rm eff}$. $H_{11}$ and $H_{22}$ contain all terms in the Hamiltonian that do not flip
$\eta_i,\eta_{i+1},\sigma_i$ or $\sigma_{i+1}$. The leading terms are $-K_i\sigma_i^z\sigma_{i+1}^z -J_i\eta_i^z\eta_{i+1}^z
-J_i\sigma_i^z\sigma_{i+1}^z\eta_i^z\eta_{i+1}^z$. In contrast, $H_{12}$ and $H_{21}$ consist of the terms that flip
$\eta_i,\eta_{i+1},\sigma_i$ and/or $\sigma_{i+1}$. This gives
\begin{eqnarray}
H_{12}= P_1 [&-&h_i\sigma_i^x -h_{i+1}\sigma_{i+1}^x -g_i \eta_i^x -g_{i+1}\eta_{i+1}^x \nonumber \\
             &-&h_i\eta_i^x\sigma_i^x - h_{i+1} \eta_{i+1}^x\sigma_{i+1}^x ]P_2 ~.
\label{eq:H12d}
\end{eqnarray}
Inserting this into the effective Hamiltonian (\ref{eq:projected}) as before yields the transverse field $h_{\rm eff}$
and two-spin field $g_{\rm eff}$ acting on the cluster variables $\sigma_{\rm eff}$ and $\eta_{\rm eff}$. Their values are
given in (\ref{eq:SDRG_bond}).

Note that the SDRG steps (c) and (d) are identical  to the site and bond decimations employed in the weak-coupling
($\epsilon<1$) analysis of Refs.\ \cite{CarlonLajkoIgloi01,GoswamiSchwabChakravarty08}.

\section{Multicritical point}
\label{appendix:MCP}

The multicritical point separating the two-phase and three-phase regions is located at $\delta=0, \epsilon=1$.
In this appendix, we sketch the derivation of the SDRG flow on the self-duality line $\delta=0$ for $\epsilon>1$
but close to the multicritical point.

Let us begin with a qualitative discussion.
For $\epsilon \approx 1$ and strong disorder, initially almost all SDRG steps are site decimations (c) or
bond decimations (d), thus the RG flow is identical to that of the random transverse-field Ising chain.
Under these steps, $\epsilon$ increases rapidly. When the typical $\ln\epsilon$ reaches the
width of the $\ln J$ and $\ln h$ distributions, the character of the SDRG flow changes. Now, most steps are
``mixed'' decimations of types (a) and (b). Under these steps, the magnetization rapidly drops to zero while
the polarization (product order parameter) stops decreasing and reaches a nonzero asymptotic value. Thus, the
RG scale at which $\ln\epsilon$ reaches the width of the $\ln J$ and $\ln h$ distributions determines the
correlation length and the polarization.

For a quantitative analysis of this SDRG flow, we start from the recursion relations for the coupling strengths
$\epsilon_{J,i}$ and $\epsilon_{h,i}$ defined by (\ref{eq:SDRG_site}) and (\ref{eq:SDRG_bond}). Written in terms
of logarithms, they read
\begin{eqnarray}
\ln \tilde\epsilon_h &=& \ln \epsilon_{h,i}+\ln\epsilon_{h,i+1} +\ln[(1+\epsilon_{J,i})/2]
\label{eq:epsilon_h_recursions}\\
\ln \tilde\epsilon_J &=& \ln \epsilon_{J,i-1}+\ln\epsilon_{J,i} +\ln[(1+\epsilon_{h,i})/2]~.
\label{eq:epsilon_J_recursions}
\end{eqnarray}
We follow the $\epsilon$-flow from $\ln\epsilon \ll 1$ to $\ln\epsilon \sim  P_0^{-1}, R_0^{-1}$ where $P_0$ and $R_0$
are the inverse widths of the $\ln J$ and $\ln h$ distributions. Two regimes need to be distinguished, $\ln \epsilon <1 $
and $\ln\epsilon >1$.

For $\ln \epsilon <1$, we expand in $\delta^{(\epsilon)} = \ln\epsilon \approx \epsilon-1$, and  equations
(\ref{eq:epsilon_h_recursions}) and (\ref{eq:epsilon_J_recursions}) simplify to
\begin{eqnarray}
\tilde \delta^{(\epsilon)}_h &=& \delta^{(\epsilon)}_{h,i} + \delta^{(\epsilon)}_{h,i+1} + \frac 1 2 \delta^{(\epsilon)}_{J,i}~,
\label{eq:delta_h_recursions}\\
\tilde \delta^{(\epsilon)}_J &=& \delta^{(\epsilon)}_{J,i-1} + \delta^{(\epsilon)}_{J,i} + \frac 1 2 \delta^{(\epsilon)}_{h,i}~.
\label{eq:delta_J_recursions}
\end{eqnarray}
The recursions can be understood as special cases of the general recursion $\tilde x_i = x_{i-1} +x_{i+1} + Y x_i$ with $Y=1/2$
\footnote{$x_i$ comprises both site and bond $\epsilon$, arranged in an alternating fashion.}.
The flow of variables governed by such recursions close to the infinite-randomness fixed point (of the random transverse-field
Ising chain) was studied in detail by
Fisher \cite{Fisher94}. He found that the typical $x$ scales like $\Gamma^{\phi_Y} =[\ln(\Omega_I/\Omega)]^{\phi_Y}$
with decreasing SDRG energy scale $\Omega$. The exponent $\phi_Y$ is given by $\phi_Y =[1+(5+4Y)^{1/2}]/2$.
(In contrast to (\ref{eq:width1}), we need to use the ``asymmetric'' version of Fisher's results because all our $\delta^{(\epsilon)} > 0$.)
Thus, in the first regime ($\ln \epsilon <1$), the typical $\ln \epsilon$ scales as
\begin{equation}
\ln\epsilon_{\rm typ} \approx \Gamma^{\phi_{1/2}} \,\ln \epsilon_0 ~, \qquad \phi_{1/2}=\frac 1 2 \left(1+\sqrt{7}\right)
\label{eq:eps_scaling_1}
\end{equation}

\begin{figure}[t]
\includegraphics[width=8.5cm]{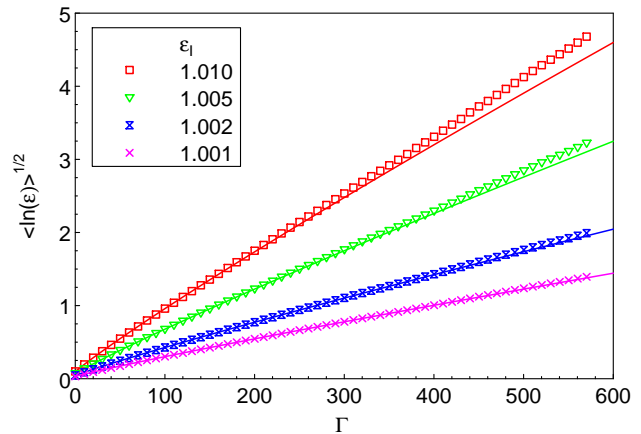}
\caption{(Color online) $\langle \ln(\epsilon) \rangle^{1/2}$ vs.\ $\Gamma$ for four different systems on the self-duality line
$h_{\rm typ} = J_{\rm typ}$ close to the multicritical point ($w=8$ and $\epsilon_I=1.001, 1.002, 1.005$ and 1.01).
Each curve stems from a single long chain of $2.5\times 10^7$ sites. The solid lines are fits of the data
in the range $\langle\ln\epsilon\rangle<0.5$ to
$\langle\ln\epsilon\rangle = C (\Gamma-\Gamma_0)^{\phi_{1/2}}$ with $\phi_{1/2} \approx 1.823$, see eq.\ (\ref{eq:eps_scaling_1}).}
\label{fig:epsilon_gamma}
\end{figure}

In the second regime, $\ln \epsilon >1$, we can approximate the recursions (\ref{eq:epsilon_h_recursions}) and (\ref{eq:epsilon_J_recursions})
for $\delta^{(\epsilon)} = \ln \epsilon$ by
\begin{eqnarray}
\tilde \delta^{(\epsilon)}_h &=& \delta^{(\epsilon)}_{h,i} + \delta^{(\epsilon)}_{h,i+1} +  \delta^{(\epsilon)}_{J,i}~,
\label{eq:delta_h_recursions_2}\\
\tilde \delta^{(\epsilon)}_J &=& \delta^{(\epsilon)}_{J,i-1} + \delta^{(\epsilon)}_{J,i} +  \delta^{(\epsilon)}_{h,i}~.
\label{eq:delta_J_recursions_2}
\end{eqnarray}
These recursions are of the same type as (\ref{eq:delta_h_recursions}) and (\ref{eq:delta_J_recursions}), but with $Y=1$.
Thus, in the second regime, $\ln \epsilon$ scales as
\begin{equation}
\ln\epsilon_{\rm typ} \sim \Gamma^{\phi_{1}}  ~, \qquad \phi_{1}=2~.
\label{eq:eps_scaling_2}
\end{equation}
To test the predictions (\ref{eq:eps_scaling_1}) and (\ref{eq:eps_scaling_2}), we implemented the strong-disorder renormalization
group numerically. Figure \ref{fig:epsilon_gamma} shows $(\ln\epsilon_{\rm typ})^{1/2}$ as a function of $\Gamma$ for systems located on the
self-duality line $h_{\rm typ} = J_{\rm typ}$. We employed identical
power-law distributions $P_I(J)=J^{-1+1/w}/w$ and $R_I(h)=h^{-1+1/w}/w$ for the interactions ($0<J<1$)
and transverse fields ($0<h<1$), with $w$ being a measure of the disorder. The coupling strength is uniform and close to the
multicritical value $\epsilon_I=1$.
The figure shows that the data in the range $\langle \ln(\epsilon) \rangle > 1$ lie on straight lines, i.e., they follow
(\ref{eq:eps_scaling_2}) as predicted. For $\langle \ln(\epsilon) \rangle < 1$ the data curve downward suggesting a smaller exponent.
In fact, the data in the range $0<\langle \ln(\epsilon) \rangle<0.5$ can be very well fitted with functions of the form
$\langle\ln\epsilon\rangle = C (\Gamma-\Gamma_0)^{\phi_{1/2}}$, in agreement with (\ref{eq:eps_scaling_1}).

Let us now combine the two regimes. We consider a (bare) system close to the multicritical point, $0<\ln \epsilon_I \ll 1$,
with strong initial disorder, i.e., the widths of the bare distributions
of $\ln J$ and $\ln h$ are large, $P_I^{-1} = R_I^{-1} \gg 1$. Under repeated site and bond decimations (SDRG steps c and d),
$P_0^{-1}=R_0^{-1} = P_I^{-1} \Gamma$ while the typical $\ln\epsilon$ scales as $\ln\epsilon \sim \Gamma^2 (\ln \epsilon_I )^{2/\phi_{1/2}}$
once $\ln \epsilon >1$. Setting $\ln \epsilon = P_0^{-1}$ gives the crossover SDRG scale
\begin{equation}
\Gamma_x = \frac 1 {P_I (\ln\epsilon_I)^{2/\phi_{1/2}}}~.
\label{eq:Gamma_x}
\end{equation}
The correlation length is given by the length scale corresponding to $\Gamma_x$,
\begin{equation}
\xi \sim \ell_x \sim \Gamma_x^2 \sim (\ln\epsilon_I)^{-4/\phi_{1/2}} \approx (\epsilon_I-1)^{-4/\phi_{1/2}}~.
\label{eq:nu_epsilon}
\end{equation}
The correlation length exponent $\nu_\epsilon$ thus takes the value $\nu_\epsilon=4/\phi_{1/2}$ as given in
(\ref{eq:MCP_eps}). The product order parameter (polarization) $M_p$ can be found by noting that $\sigma$-clusters are not
decimated anymore once $\Gamma > \Gamma_x$. $M_p$ is thus given by its value at $\Gamma_x$.
\begin{equation}
M_p = n(\Gamma_x)\, \mu(\Gamma_x) \sim \Gamma_x^{-2+\phi_0} \sim (\ln\epsilon_I)^{2(2-\phi_0)/\phi_{1/2}}
\label{eq:beta_epsilon}
\end{equation}
where $n(\Gamma)$ and $\mu(\Gamma)$ are the number and moment of clusters surviving at SDRG scale $\Gamma$.
Using  $\phi_0=[1+\sqrt{5}]/2$ yields the order parameter exponent $\beta_\epsilon$ given in (\ref{eq:MCP_eps}).

\bibliographystyle{apsrev4-1}
\bibliography{../00Bibtex/rareregions}

\end{document}